\documentstyle[amssymb,aps,pre,multicol,epsfig]{revtex}

\begin{document}
\title{Dynamical models for sand ripples beneath surface waves}

\author{Ken Haste Andersen$^{1,2,3}$, Marie-Line Chabanol$^{1,4}$ and
  Martin van Hecke$^{1,5}$}

\address{$^1$ Center for Chaos and Turbulence Studies, The Niels Bohr
  Institute, Blegdamsvej 17, 2100 Copenhagen \O, Denmark.\\ $^2$
  Institute of Hydraulic Research and Water Resources, the Technical
  University of Denmark, 2800 Lyngby, Denmark.\\ $^3$ Dipartimento di
  Fisica, Università degli Studi di Roma La Sapienza, Piazzale Aldo
  Moro 2, I-00185 Roma, Italy. \\
  $^4$ Institut Fourier, 100 rue des Math\'ematiques, Domaine
  Universitaire, 38400 Saint Martin d'H\`eres, France.\\ $^5$
  Max-Planck-Institut f\"ur Physik komplexer Systeme, N\"othnitzer
  Str. 38 01187 Dresden, Germany.}

\date{\today}
\maketitle
\begin{abstract}
  We introduce order parameter models for describing the dynamics of
  sand ripple patterns under oscillatory flow. A crucial ingredient of
  these models is the mass transport between adjacent ripples, which
  we obtain from detailed numerical simulations for a range of ripple
  sizes. Using this mass transport function, our models predict the
  existence of a stable band of wavenumbers limited by secondary
  instabilities. Small ripples coarsen in our models and this process
  leads to a sharply selected final wavenumber, in agreement with
  experimental observations.
\end{abstract}


\begin{multicols}{2}  

\section{Introduction}
\begin{figure}
  \noindent
  \begin{minipage}{86mm}
    \begin{center}
    \epsfig{file=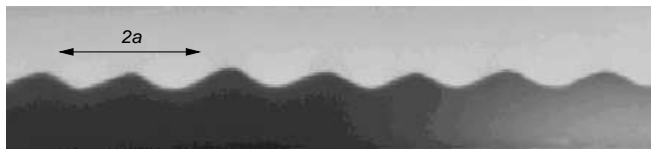,width=86mm}
    \vspace{1mm} 
    \caption{Side-view of a vortex ripple pattern under oscillatory
      flow in a long, slender channel (courtesy of J.L. Hansen).  The
      arrow indicates the amplitude of the fluid oscillations in the
      bulk.  }
    \label{fig:photo}
    \end{center}
  \end{minipage}
\end{figure}

When a flat surface of sand is exposed to the flow of air or water,
{\em patterns} known as ripples, dunes, sandwaves and draas are formed
\cite{Bagbook,raud:97,Bag,ourexp,ande:99,nish:98}.  Here we focus on
the so-called {\em vortex ripples} \cite{Bagbook}
(Fig.~\ref{fig:photo}) which are created by oscillatory fluid flow,
e.g., beneath surface waves. Ripples are of interest to coastal engineers
since their properties determine the friction of the flow in the
coastal region \cite{lofq:80,slea:82,math:96,fred:99}, the dissipation
of surface waves \cite{ande:00} and the net sediment transport over
the ripples \cite{hans:94}. More recently ripples have attracted the
attention of physicists interested in non-equilibrium systems
\cite{raud:97,ourexp,anderson,nish:93,prig:98,terz:98,Csa,Bet,Steg,Scherer}.

The physics underlying sand ripple formation involves the interaction
between the turbulent fluid flow and a granular medium, and is
therefore extremely complex. A description of the pattern forming
aspects is hindered by the strong nonlinearity of the fully developed
ripples due to the subcritical nature of the initial bifurcation from
a flat bed. Previous theoretical studies of this initial bifurcation
\cite{Blon,Blon2,Vitt,foti:95} have described the {\em onset} of
ripple formation. Vortex ripple pattern formation occurs, however, far
from equilibrium: typical wavelengths of fully developed ripples can
be a factor five larger than those predicted by (weakly nonlinear)
analysis \cite{foti:95}.

In this paper we will discuss the {\em pattern forming} aspects of
fully developed vortex ripples.  Many of the problems associated with
the complicated underlying phenomenology can be circumvented by noting
that the {\em sizes} of the ripples are the most relevant parameter
for determination of their dynamics; further details of their shapes
are not important. Dynamical equations for the evolution of the
ripples can then be constructed once the mass exchange between ripples
of certain sizes is known. We base our expression for this mass
exchange on detailed numerical simulations of the flow and sand
transport over vortex ripples (see below), hence going beyond a pure
``toy-model'' approach.  As far as we are aware, the model presented
here is the first to capture both

\noindent instabilities and coarsening of fully developed vortex
ripples.

The outline of the paper is as follows. We start with a brief
description of the main phenomenology of ripples in section \ref{vr}.
Although the amplitude of the fluid oscillations determines the length
of the ripples, a dimensional analysis (section \ref{dimanalsec})
reveals that the most relevant dimensionless control parameter is the
Shields parameter which characterizes stress at the sandy surface. We
discuss our numerical simulations of the mass exchange between vortex
ripples in section \ref{nums}. Section \ref{1Dmodel} is devoted to the
formulation of simple ripple models in one-dimensional geometries.
The linear stability of these models is performed in section
\ref{sec:stability}, and the coarsening and selection of the final
ripple patterns starting from random initial conditions is discussed
in section \ref{num}. In section \ref{2Dmodel} we extend our model to
two dimensions and discuss the impact of defect motion on the
selection of the final two--dimensional pattern.

\section{Vortex Ripples}\label{vr}

Following the much earlier work of Ayrton \cite{Ayr}, the study of
vortex ripples was taken up again by Bagnold in 1946 \cite{Bag}. In
this seminal study, Bagnold distinguished between {\em rolling grain
  ripples} and {\em vortex ripples}. The former are generated when
starting from an unstable flat bed \cite{ande:99} and consist of small
triangular ridges separated by a comparatively long stretch of flat
bed. These rolling grain ripples grow and coarsen to become vortex
ripples with no flat bed between them.  Here the flow is dominated by
{\em separation bubbles} (vortices) on the lee sides of the fully
developed ripples.
% Vortex ripples under
%oscillatory flow are, on average, symmetrical and do not
%migrate.
We will concentrate on these fully developed vortex ripples, since
recent studies have confirmed \cite{Steg,Scherer} that rolling grain
ripples essentially constitute a transient.
%that occurs on the way from a flat
%bed to the formation of vortex ripples.

Many experiments have studied the average wave length of fully
developed ripples \cite{ding:75,niel:81,wibe:94,tray:99} as a function
of, e.g., the amplitude and frequency of the fluid motion.  It appears
that the (dimensional) length of the ripples ($\lambda_{dim}$) is
proportional to the amplitude of the oscillatory flow ($a$), and
roughly independent of its frequency. Estimates in the literature of
the proportionality constant $\lambda_{dim}/a$ range from one to two,
with a preference for values around 1.3 (see Fig.~8 in
\cite{niel:81}).

Recently the ripples have also been studied from the view point of
pattern formation. Both Scherer {\em et al.} \cite{Scherer} and
Stegner and Wesfreid \cite{Steg} studied a one--dimensional annular
system in which the conservation of sand is guaranteed. Stegner and
Wesfried \cite{Steg} observed strong hysteresis when the driving
amplitude of fully developed ripples was ramped up and down: an
increase in the amplitude of the driving yielded larger ripples, while
for a decrease, the ripples did not change length. Lofquist also
\cite{lofq:80} observed hysteretic behavior, but in this case the
ripples were initially stable for both an increase or decrease of the
driving amplitude \cite{lofq:80}. Hysteresis of the ripples was also
observed in a recent set of field measurements \cite{tray:99}.

%To summarize: vortex ripple patterns combine a number of features, of
%which the most important are ({\em i}) there is no intrinsic
%length-scale, but instead the scales are set by the driving amplitude
%({\em ii}) the initial bifurcation is strongly subcritical ({\em iii})
%the fully developed pattern show hysteresis.

\subsection{Dimensional analysis and setup of the problem}\label{dimanalsec}

Ripples are governed by a large number of dimensional parameters which
characterize fluid flow and sand. We will show that while in
general three dimensionless parameters (density ratio of fluid and
sand grains $s$, settling velocity $w_s$ and maximum {\em Shields
parameter} $\theta_{max}$) characterize the system, for the case of
interest here (sand/water systems in the regime where suspension is
unimportant) the only free parameter is the Shields parameter.

Ripple formation is driven by an oscillatory fluid motion with
amplitude $a$ and angular frequency $\omega$.  The Reynolds number
$Re$ for this situation is $a^2 \omega/\nu$, where $\nu$ is the fluid
viscosity. For water in a typical experimental situation
($a\!=\!5$~cm, $\omega\!  =\!3$~s$^{-1}$), $Re$ is of order $10^3$,
and the flow is turbulent. Therefore, large scale flow structures such
as separation bubbles are independent of the Reynolds number and
hence viscosity. For turbulent flow, the roughness of the bed is of
minor importance as long as the typical grain sizes are much smaller
than $a$. The only relevant length scale is then $a$, which we use to
define the non-dimensional ripple length as
$\lambda\!=\!\lambda_{dim}/a$.  The large scale flow is then
completely specified by the boundary conditions, i.e., the shape of
the ripples.

The sand introduces four new dimensional parameters into the
problem. These are, respectively, the density of water $\rho_w$ and
sand $\rho_s$, the median diameter of the grains $d$ and gravity $g$.
From these we form the following three non-dimensional parameters:
\begin{equation}
  s = \frac{\rho_s}{\rho_w},\quad w_s =
  \frac{w_{s.dim}}{a\omega},\quad \theta = \frac{\tau_{bed}}{\rho_w
    (s-1)gd},
\end{equation}

\end{multicols}
\begin{figure}[htbp]
  % .cps from groft/sharp/e401.  % lower plots from groft/sharp/e401.m.
  \begin{center} 
    \hspace*{-1.2cm} \epsfig{file=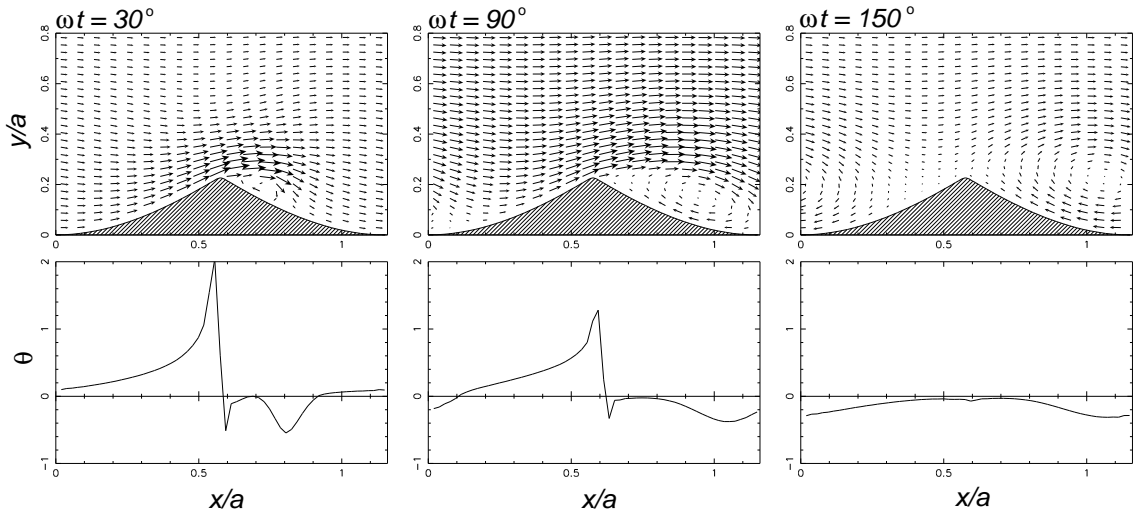, width=15cm}
\vspace{3mm}
    \caption{The flow over a ripple at three instants in the first
      half of the wave period. The system consists of a single ripple
      in a system with periodic boundary conditions and length  $\lambda=1.15$.
      The bottom three figures show the spatial profile of the Shields
      parameter on the bed at the corresponding times. For this case
      $\theta_{max} = 0.13$.}
    \label{fig:flow} 
  \end{center}
\end{figure}
\begin{multicols}{2}
The relative density of the grains $s$ has a value of $2.65$ for
quartz sand in water. The settling velocity $\omega_s$ characterizes
the amount of sand kept in suspension; here we assume a regime where
the settling velocity is large ($w_s\!\gtrsim\!0.15$) such that
suspension is not important.  This leaves us with the last parameter,
$\theta$,

\noindent which is known as the {\em Shields parameter}. The Shields
parameter expresses the ratio between the drag and gravitational
forces on a single grain and depends on the shear stress
$\tau_{bed}(x,t)$, which varies with time and along the profile of the
ripple. Following \cite{niel:79} we propose to use the maximum shear
stress on a {\em flat} bed, $\tau_{max}$, to characterize the flow. For
laminar flow $\tau_{max}$ can be found exactly from the solution of
Stokes' second problem \cite{ll}.
%$  \tau_{max} = a\omega^{3/2}\sqrt{\nu}\rho_w.$
For turbulent conditions, which prevail here, an analytical expression
does not exist. We will follow coastal engineers in using an empirical
relation for the maximum shear stress \cite{fred:92}:
\begin{equation}
  \tau_{max} =
  0.02\rho_w\left(\frac{a}{k_N}\right)^{-0.25}(a\omega)^2.
  \label{eq:fw}
\end{equation}
Note that the instantaneous Shields parameter on a {\em rippled} bed
$\theta(x,t)$ can be several times larger than $\theta_{max}$.

The transport of sand $q$ takes place in a thin layer above the bed,
the so-called {\em bed load} layer (for an introduction to sediment
transport see chapter 7 in ref.~\cite{fred:92}). The
non-dimensionalized flux of sand $\phi\!\equiv\!q/\sqrt{g(s-1)d^3}$ in
the bed load layer is a function of the local Shields parameter and
can be modeled as:
\begin{equation}
  \phi = \alpha (\theta - \theta_c)^\beta. \label{equ:qb}
\end{equation}
When $\theta(x,t)$ smaller than a critical value $\theta_c$ for all
$x$, which for turbulent boundary layers is approximately $0.06$
\cite{fred:92}, sand grains do not move and the ripple profile
freezes. The constants $\alpha$ and $\beta$ have been determined
empirically by Meyer-Peter and M\"uller \cite{meye:48} to be
approximately $\alpha\!=\!8$ and $\beta\!=\!1.5$, which are in good
agreement with theoretical estimates \cite{enge:76,kova:94}. The
formation and the dynamical properties of the ripples are mainly
determined by the fluid flow, so the exact values of the constants
$\alpha$ and $\beta$ together with the detailed form of
Eq.~(\ref{equ:qb}), turn out to be relatively unimportant for the
content of this work.

%In conclusion, for ripples made of sand and when suspension is not
%important, the system is characterized by the maximum shear stress
%$\theta_{max}$ and the non-dimensionalized ripple profile.

\subsection{Numerical studies and mass transport}\label{nums}
The computational model that we have developed to study the ripples
calculates turbulent fluid flow over ripples based on the standard
$k$-$\omega$ turbulence model \cite{wilc:88,kensthesis}.  Once this
flow is known, the sediment transport, which is governed by the shear
stress on the bed, can be calculated from Eq. (\ref{equ:qb}).

In Fig.~\ref{fig:flow} we show some results for the flow and the
non-dimensionalized shear stress $\phi$ for $\lambda\!=\!1.15$. We see
that 
\begin{figure}[tb]
  \noindent
  \begin{minipage}{86mm}
    \begin{center}
        \vspace{-1.2cm}
    \epsfig{file=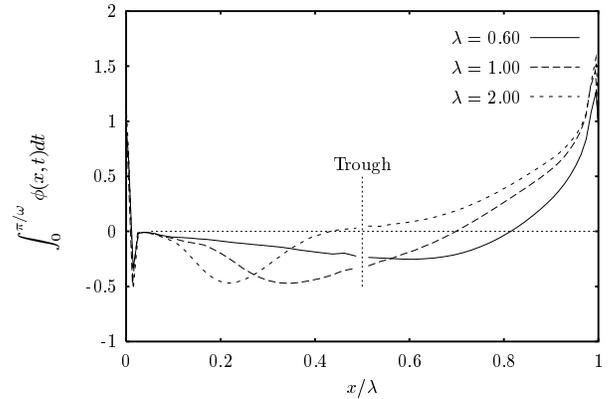, width=10.5cm}
    \vspace{1mm}
    \caption{The spatial profile of the net mass transport in
      the first half of the wave period ($\int_0^{\pi/\omega}
      \phi(x,t)dt$) for ripples of length $\lambda\!=\!0.6,\ 1.0$ and
      2.0. Note that the ripple trough is located in the middle of the
      figure.  ($\theta_{max} = 0.13$).}
    \label{fig:three}
    \end{center}
  \end{minipage}
\end{figure}
\noindent there are two mechanisms that generate the shear stress on
the bed, namely the converging flow on the ``wind'' side of the ripple
(here left) and the separation bubble formed in the lee (right)
side. Typically, these stresses are several times stronger than the
stresses on a flat bed. The separation bubble, where the flow near the
bed is directed opposite to the mean flow direction, is clearly
visible. This bubble moves out into the trough of the ripple ($\omega
t\!=\!90^\circ$), where it stays ($\omega t\!=\!150^\circ$) until it
is thrown over the crest as the flow reverses.

The shear stresses are uphill on both sides of the ripple, and
consequently sand is transported from the trough toward the crest; the
result is a steepening of the ripple profile. This steepening
continues until the slopes of the ripple reach the angle of repose,
when avalanches limit the growth of the ripple slopes. As a
consequence, most slopes of the fully developed ripples are close
to the angle of repose. These fully developed ripples are thus
approximately triangular, joined by smooth troughs, which is also
evident from experiments \cite{Steg}.

Ripples interact by exchanging sand between their neighbours over the
troughs. The amount of this mass flow is closely connected to the
extension and strength of the separation bubble. We have studied this
mass transport as a function of the non-dimensional ripple size
$\lambda$.  In Fig.~\ref{fig:three} we show the net sediment transport
during the first half wave period, for short ($\lambda\!=\!0.6$),
medium ($\lambda\!=\!1.0$) and large ripples ($\lambda\!=\!2.0$).  For
short ripples the separation bubble almost covers the space between
the two ripple crests, but it is not very strong, giving rise to a
small transport. For long ripples, the separation bubble does not
reach over the trough, again giving only a small mass exchange between
adjacent ripples.  Most mass is exchanged for medium sized ripples,
where the separation bubble is both strong and reaches over the
trough.

We define $f$ as the amount of sand transported over the
trough during the first half wave period:

\begin{figure}[tbp]
  % from groft/sharp/e500.gnu \noindent 
  \noindent
  \begin{minipage}{86mm} 
    \epsfig{file=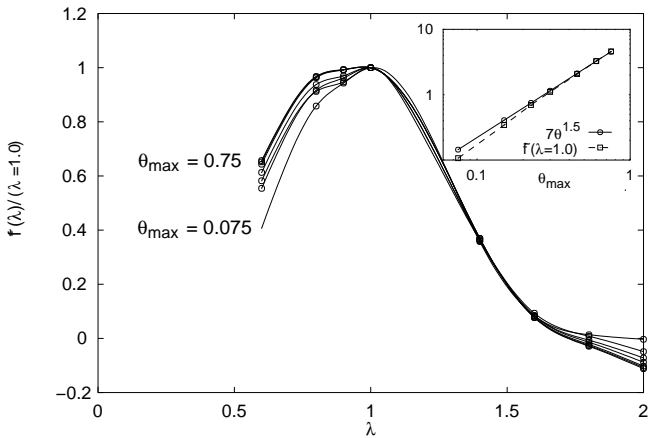, width=86mm} \vspace{1mm}
    \caption{$f(\lambda)$ for $\theta_{max} = 0.075$, 0.15, 0.23,
      0.30, 0.45, 0.60 and 0.75, scaled with $f(\lambda=1.0)$. The
      inset show $7\theta_{max}^{1.5}$ (full line) and
      $f(\lambda=1.0)$ (dashed line) in double log scaling.}
    \label{fig:interaction} \end{minipage}
\end{figure}

\begin{equation}
  f(\lambda) = -\int_0^{\pi/\omega} \phi(x_{tr},t) dt,
  \label{equ:ftilde}
\end{equation}
where $x_{tr}$ is the position of the trough.  The minus sign is
simply related to the fact that the fluid and mass flows have opposite
directions during each half period; here we wish to have a positive
$f(\lambda)$.

The rescaled mass exchange $f(\lambda)/ f(\lambda=1.0)$ is shown in
Fig.~\ref{fig:interaction} for values of $\theta_{max}$ ranging from
$0.075$ to $0.75$.  The rescaled graphs of the mass exchange
{\em{(i)}} collapse in good approximation, and {\em{(ii)}} have a
single maximum around $\lambda\!=\!1.0$. In our model, developed in
section \ref{1Dmodel} below, we will incorporate these two properties.
If the critical Shields parameter had been zero, the rescaling factor
$f(\lambda=1.0)$ would have been proportional to $\theta_{max}^{1.5}$.
That this is almost, but not exactly, the case is seen in the inset in
Fig.~\ref{fig:interaction}.

\section{Discrete models for one-dimensional ripples}\label{1Dmodel}

\begin{figure}[htbp]
  \noindent \begin{minipage}{86mm} \begin{center}
  \epsfig{file=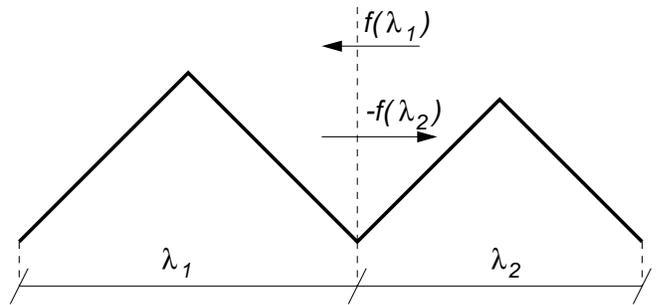, width=8.6cm} \vspace{2mm} \caption{An
  example of how the transport in the trough between two ripples is
  constructed as the transport in the two half periods.}
  \label{fig:example1} \end{center} \end{minipage}
\end{figure}

In this section we will introduce and study simple models for ripples
in one-dimensional geometries. We assume that the angles of the ripple
slopes are fixed, so that the only degrees of freedom are the lengths
of their left and right slopes.  Two different versions of the model
will be described. In the simplest case we only take the total ripple
sizes $\lambda_i$ into account (see Fig.~\ref{fig:example1}). The
ensuing ``minimal model'' is formulated in section \ref{quali}, and is
analyzed theoretically in section \ref{sec:stability}.  A more refined
model which takes the lengths of left and right slopes into account is
presented in section \ref{1Dmodel:formu} (see Fig.~\ref{fig:sketch}),
and numerical simulations of this model are presented in section
\ref{num}.

\subsection{Minimal model}\label{quali}
In this model ripples are triangular and symmetric and characterized
by their length $\lambda_i$. We will now determine the mass transfer
between two ripples with lengths $\lambda_1$ and $\lambda_2$
(Fig.~\ref{fig:example1}) from the information that we have for the
mass transfer between equal ripples. When $\lambda_1$ and $\lambda_2$
are approximately equal, one expects the size and strength of the
separation bubble emanating from the crest of ripple 1 to be
independent of the size of ripple 2. This is our central assumption:
{\em the mass transport during a half period only depends on the size
of the ripple that creates the separation bubble.}

Let us denote the first half period of the driving, when the flow is
from left to right, by a subscript $I$, and the second half by $II$.
Under our assumption stated above, we obtain: $\Delta
m_I\!=\!f(\lambda_1)$ and $\Delta m_{II}\!=\!-
f(\lambda_2)$, where $\Delta m_I$ denotes the change in the
mass of ripple 1 in the first half period. During each half period,
the amount of mass transported is small in comparison to the mass of a
single ripple. We therefore can neglect changes in the ripple shapes
during a half period, and obtain the mass flow during a full period,
$\Delta m$, by simply adding up the half period mass--flows:
\begin{equation} \label{simpleeq}
  \Delta m = f(\lambda_1)- f (\lambda_2)~.
\end{equation}
Clearly Eq. (\ref{simpleeq}) can be extended to the case of a row of
ripples. Then the mass-flow to ripple $i$, $\Delta m_{i}$, is due to
interactions with both ripple $i-1$ and $i+1$: $ \Delta m_{i}\!=\!2
f(\lambda_i)- f (\lambda_{i+1}) - f (\lambda_{i-1}) $.

To close the equations we need to relate the mass flow to a change in
the size of the ripples. Since the mass-flow is small, it is
reasonable to assume that the change in ripple size is linear in the
mass transport. The greatest simplification is obtained if we assume
all ripples to be of near equal size, so that the ratio $\Delta m
/\Delta \lambda$ is equal for all ripples. Taking the continuum time
limit and rescaling time to absorb a proportionality constant we
obtain:
\begin{equation}\label{bonehead}
  d \lambda_i/dt = -f(\lambda_{i-1}) +2 f(\lambda_i) -
  f (\lambda_{i+1})~.
\end{equation}
The total length of a system of ripples evolving according to
Eq.~(\ref{bonehead}) is conserved, but the total mass is not; we will
discuss this further in section \ref{1Dmodel:formu}.

Finally, we supplement the model with an annihilation rule which
removes ripples that have shrunk to size zero, and a creation rule,
which adds ripples in the troughs between ripples of sizes larger than
a certain length $\lambda_{max}$ that will be specified in the next
section.

\subsection{Equilibria and stability}
\label{sec:stability}
There are three types of equilibria in the minimal model
(\ref{bonehead}). {\em{(i)}} Homogeneous states where all $\lambda$'s
are equal.  {\em{(ii)}} ``Period two'' states, for which
$f(\lambda_i)\!=\!f(\lambda_{i+1})$ but $\lambda_i \neq \lambda_{i+1}$
(see Fig.~16 in \cite{Scherer} for similar states). {\em{(iii)}} More
complicated equilibria constructed by arbitrary juxtapositions of
ripples of lengths $\lambda_a$ or $\lambda_{b}$ when
$f(\lambda_a)\!=\!f(\lambda_b)$.

The linear stability of the homogeneous state follows from setting
 $\lambda_{i}\!=\!\lambda_{eq} + \delta_i$ and linearizing
 Eq.~(\ref{bonehead}):
\begin{equation}
  d \delta_i / d t = -f'(\lambda_{eq})(\delta_{i-1}-2\delta_i
  +\delta_{i+1})~.
\end{equation}
This is the linear stability equation for the space-discretized
diffusion equation, with diffusion coefficient $-f'(\lambda_{eq})$,
and the sign of $f'$ will be important. As we demonstrated in section
\ref{nums}, the mass transport $f$ displays a single maximum as a
function of $\lambda$ at a value that we will refer to as
$\lambda_{min}$. When $\lambda_{eq}$ is larger (smaller) than
$\lambda_{min}$, $-f'(\lambda_{eq})$ is positive (negative) and the
pattern is stable (unstable). Hence the smallest possible stable
wavelength is $\lambda_{min}$, where $f$ has a maximum. This
instability can be seen directly from the mass transport: when we inspect
two unequal adjacent ripples with sizes larger than $\lambda_{min}$ we
obtain from Eq. (\ref{simpleeq}) that mass will flow from the larger
to the smaller ripple, hence leading to a stable equilibrium, while if
their sizes are smaller than $\lambda_{min}$ mass flows from the
smaller to the larger ripple, leading to an instability.

An additional instability occurs for {\em large} ripples when their
troughs lie outside the separation zone (see $\lambda\!=\!2.0$ in
Fig.~\ref{fig:three}); in this case the flow creates new ripples in
the troughs.  This instability has been observed in experiments
\cite{ourexp,lofq:80} and also in our numerical studies
\cite{kensthesis}. This instability is consistent with our models when
we assume that $f$ is defined for arbitrary small ripples.
For a homogeneous pattern of large ripples where
$f(\lambda_{eq}) < f(0)$, infinitesimal ripples
inserted between the large ripples will gain mass and grow, and the
maximum value $\lambda_{max}$ where homogeneous patterns are stable,
is given by $f(\lambda_{max})\!=\!f(0)$. This is the
motivation for having a creation rule in the model.

The period-two and more complicated equilibrium states can be shown to
be unstable in our framework \cite{notep2}.
Thus our model illustrates an important consequence of the shape of
the mass exchange function. {\em There is a }\\
\begin{figure}[htbp]
  \noindent \begin{minipage}{86mm} \begin{center}
  \epsfig{file=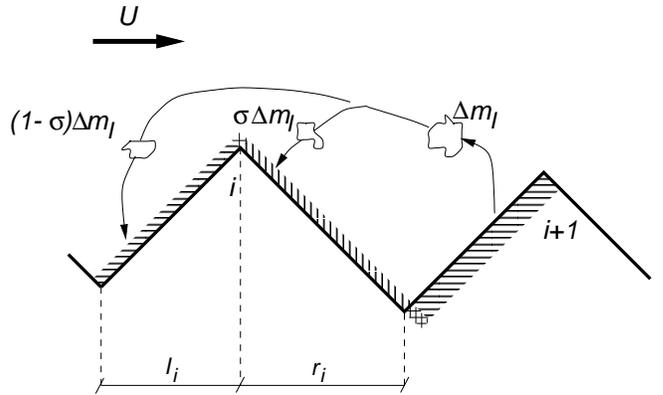, width=8.6cm} \vspace{2mm} \caption{A
  sketch of the ripple profile with the triangular ripples and the
  quantities used to describe the ripples. Note that the ripples do
  not need to be symmetric. Also shown is the exchange of sediment
  from the right ripple to the left ripple during the first half wave
  period.}  \label{fig:sketch} \end{center} \end{minipage}
\end{figure}
\noindent 

{\em band of wavelengths for
which ripple patterns are stable; outside this band, short wavelength
instabilities occur.}

\subsection{Refined model}\label{1Dmodel:formu}

Both our numerical studies and experiments \cite{Steg} frequently
display ripples that are asymmetric during their evolution (although,
on average, they are not). We extend the minimal model from the
previous section to allow for asymmetric ripples by characterizing the
ripples by the length of both their left ($l_i$) and right ($r_i$)
slopes; obviously $\lambda_i\!=\!l_i + r_i$ (see
Fig.~\ref{fig:sketch}). In addition, such a model can be tuned so as
to conserve mass.
%Since we now
%will phrase all our dynamics in terms of the lenght of just one side
%of the ripples, we will use a function $f$ defined via
%$f(z)\!=\!f (2z)$.

As before we assume that the lee side of ripples determines the size
of the separation bubbles. During the first half period the bubble
takes $\Delta m_I$ mass from the left slope of ripple $i+1$, and
transports this mass to ripple $i$; the ratio between the mass
deposited on the left and right slopes of ripple $i$ is given by a
parameter $\sigma$ that we always fix at a value of $0.5$ (see
Fig.~\ref{fig:sketch}).  The mass flow in the first half period is
therefore:
\begin{eqnarray} \label{tripleI}
  \Delta m_{li.I} &=& -f(2r_{i-1}) + (1-\sigma)f(2r_i) \nonumber\\
  \Delta m_{ri.I} &=& \sigma f(2r_i) ~. \label{deltam}
\end{eqnarray}
The mass-flow in the second half period follows by symmetry. Assuming
that the mass transport is small, we can neglect the change in
ripple size during one half cycle, and add the contributions from
each half period.

To obtain a closed set of dynamical equations we have to establish how
the lengths $l_i$ and $r_i$ evolve under a certain mass flow. When an
amount of mass $M$ is deposited on the right slope of ripple $i$, we
incorporate this by an increase of $l_i$ and a subsequent decrease of
$l_{i+1}$; the length $r_i$ itself does not change.  Assuming for
simplicity the 

\begin{figure}[tbp]
  \noindent
  \begin{minipage}{86mm}
    \begin{center}
  \epsfig{file=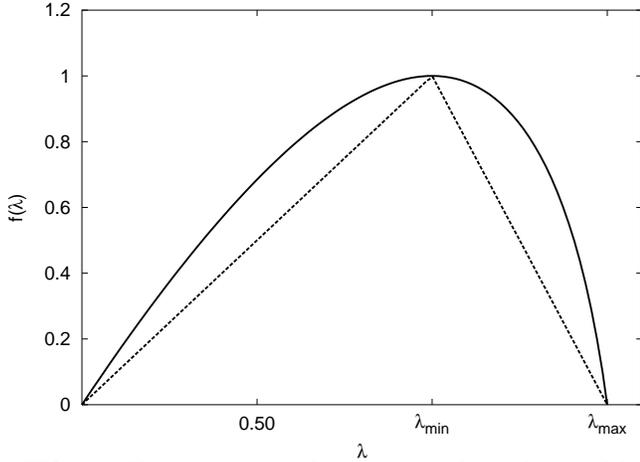, width=9cm}
  \caption{The interaction functions used in the models. Smooth
    function, (full line) and the bi-linear
    function (broken line),  for a value of 
    $\lambda_{max}\!=\!1.5$. }
  \label{fig:deltam} 
  \end{center}
  \end{minipage}
\end{figure} 

\noindent angle of repose to be $45^\circ$, we can calculate the
volume of the slab of deposited sand and find that the change in the
length of nearby ripples is:
\begin{eqnarray}
  \Delta l_{i} &=& \Delta m_{ri} /(2 r_i) \\
  \Delta l_{i+1} &= & -\Delta m_{ri} /(2 r_i)~.
  \label{changeoflength}
\end{eqnarray}

Ignoring higher order effects we obtain the total change in the
length as a function of the mass changes  as:
\begin{eqnarray}
  \Delta r_i &=& -\frac{\Delta m_{li+1}}{2 l_{i+1}} + 
  \frac{\Delta m_{li}}{2 l_{i}} \nonumber \\
  \Delta l_i &=& \frac{\Delta m_{ri}}{2 r_{i}} - 
  \frac{\Delta m_{ri-1}}{2 r_{i-1}}~. \label{refinedmodel}
\end{eqnarray} 
This relation together with the mass flow from (\ref{deltam}) defines
the refined model. This model has the same linear stability properties
as the minimal model defined in Eq. (\ref{simpleeq}).

The total length of the system is conserved, and the total mass is
approximately conserved. The masses which are ignored are associated
with the small areas that are cross-hatched in
Fig.~\ref{fig:sketch}. It is possible to formulate the model in a
strictly mass-conserving manner, by updating the slope lengths when
both removing and depositing mass, but this does not alter the model
in any substantial way.

In our numerical simulations two different forms of the mass transport
function $f(\lambda)$ were used. Both functions have a maximum at
$\lambda\!=\!\lambda_{min}=1/2$ and are zero at 0 and $\lambda_{max}$.
The simplest function that satisfies these requirements is bi-linear,
while a smooth function with a quadratic maximum that satisfies
$f(0)\!=\!f(\lambda_{max})\!=\!0$ can be constructed as the sum of a
linear function and a square root (see Fig.~\ref{fig:deltam}):
\begin{eqnarray}\label{sqrr}
  f(z) = \frac{4z}{2-\lambda_{max}} +
  \frac{\lambda_{max}(\lambda_{max}-4)}{2(\lambda_{max}-2)^2} + \nonumber \\
  \frac{\lambda_{max}}{2(\lambda_{max}-2)^2}
  \sqrt{16(\lambda_{max}-2)z + (\lambda_{max}-4)^2}~.\label{eq:f2}
\end{eqnarray}

%\begin{equation}
%  f(z) = \left\{ 
%    \begin{array}{ll}    
%  2  z & {\mbox{for}}\ z < \half \\
%      \displaystyle  \left( 1 - \frac{2 z-1}{\lambda_{max}-1}\right) 
%& {\mbox{ for }}\ z \ge \half
%    \end{array} 
%\right. ~.\label{eq:f1}
%\end{equation}

\begin{figure}[tbp]
  \noindent
  \begin{minipage}{86mm}
    \begin{center}
    \epsfig{file=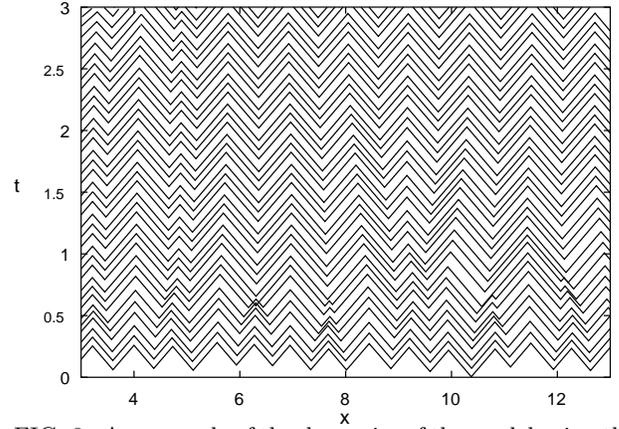, width=8cm}\\
    \caption{An example of the dynamics of the model using the linear
    interaction function and $\lambda_{max}\!=\!1.35$ The time scale
    is arbitrary.}
          \label{fig:sample} 
  \end{center} \end{minipage}
\end{figure}
\noindent This smooth function for $\lambda_{max}\!=\!1.6$ resembles
the one found from the computational flow model in section \ref{nums}.

\subsection{Coarsening of fully developed ripples}
\label{num}
When ripples are grown experimentally from a flat bed, initially
many small ripples are created. They subsequently coarsen and form a
final regular steady state with a well-defined final wavelength (see
for example Fig.~1 in \cite{Steg}). Our model shows the same behavior
for initial conditions of  (disordered) unstable small
wavelength patterns. An example of such evolution is shown in
Fig.~\ref{fig:sample}. A fast coarsening process is seen in the
beginning ($t\!<\!1$), followed by a slower relaxation toward an
equilibrium state. The important dynamical process leading to the
equilibrium state is the annihilation of ripples, with each
annihilation resulting in a longer average ripple length; creation
does not play a role here. After the final annihilation, slow
diffusive dynamics sets in.

The stability analysis performed in section \ref{sec:stability} shows
that a wide range of ripple wave lengths can be linearly stable,
namely $\lambda_{min}\!<\lambda\!<\lambda_{max}$.  We will show here
that, starting from small ripples, the dynamics leads to the selection
of a sharply defined final wavelength. We assume periodic boundary
conditions in our simulations. The parameters entering the model are
the length of the domain $L$ and the maximum ripple length
$\lambda_{max}$. The initial conditions are disordered ripples with an
average wavelength $\lambda_0\!<\!\lambda_{min}$.

We found that the final wavelength is quite independent upon the
initial average wavelength $\lambda_{0}$ (when this is sufficiently
small) and the initial degree of disorder. This result could not have
been predicted {\em a priori} from the model equations, but is in good
agreement with experimental evidence.

\begin{figure}[htb]
  \noindent \begin{minipage}{86mm} \begin{center} \epsfig{file=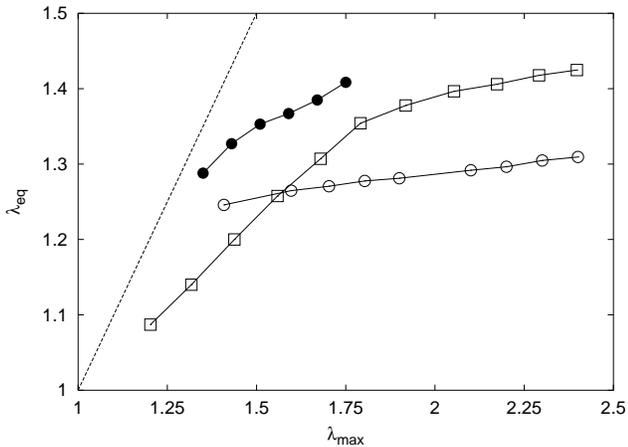,
  width=8.6cm} \vspace{2mm} \caption{The equilibrium wavelength as a
  function of $\lambda_{max}$.  The squares correspond to results
  obtained using the bi-linear interaction function, the open circles
  corresponds to the smooth function and the
  filled circles corresponds to the 2D model. The dashed line
  indicates the maximum possible wavelength,
  $\lambda\!=\!\lambda_{max}$. The initial number of ripples were
  1200, and the initial ripple length was $0.7\!\pm\!0.05$.}
  \label{leq1D} \end{center} \end{minipage}
\end{figure}

\noindent The final wavelength does, however, depend on the shape of the
interaction function and the value of $\lambda_{max}$. In
Fig.~\ref{leq1D}, $\lambda_{eq}$ is plotted as a function of
$\lambda_{max}$ for the two interaction functions. The final
wavelength appears to be a nontrivial function of $\lambda_{max}$ for
both interaction functions. The interaction function which resembles
the one from the numerical flow calculations (the smooth function with
$\lambda_{max}\!=\!1.6$) results in an equilibrium wavelength of
$\lambda_{eq}\!=\!1.28 \pm 0.03$, a result which is in good agreement
with ripple lengths measured in experiments.

%\begin{figure}[htb]
%  \noindent \begin{minipage}{86mm} \begin{center}
%  \epsfig{file=2ddiffusion-b.eps, width=6cm} \vspace{2mm} 
%\caption{A sketch of the mass
%  fluxes in the $j$-direction induced by the crests not being aligned
%  perpendicular to the flow (thick arrows).  The mass fluxes induces a
%  movement of the individual ripples and of the crests as a whole.}
%  \label{Masstransf} \end{center} \end{minipage}
%\end{figure}

\section{Two-dimensional ripple patterns }\label{2Dmodel}
Unless one forces the ripple patterns to occur in a narrow channel or
annulus, ripple patterns are two-dimensional, even though the
alignment perpendicular to the flow yields quasi one-dimensional
patterns. However, during the evolution toward the final state the
pattern contains many defects \cite{ourexp}. We have thus extended our
model to study their role in the selection of the final state.

In our two-dimensional model the individual rows of ripples are
similar to those in Fig.~\ref{fig:sketch} and are labeled by an indices
$i$ and $j$, where $j$ is the new coordinate perpendicular to the driving
direction.  In the $i$-direction, the mass flow is given by expression
(\ref{deltam}). We then determine which ripples are neighbors in the
$j$-direction, and impose an angle of repose in the $j$ direction by
inducing a flow of mass between ripple $j$ and $j+1$ when their height
difference is above a certain maximum. At a defect, such mass flow can
nucleate a new ripple in the trough of an adjacent row.  Finally, it
is reasonable to  expect that a

\begin{figure}[htb]
  \noindent
  \begin{minipage}{86mm}
    \begin{center}
  \epsfig{file=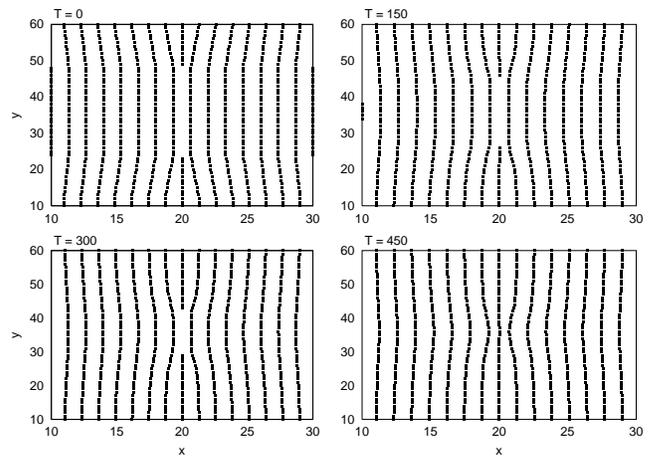, width=8.6cm}
  \caption{An example of a climbing defect in a system with 30 ripples
    in the middle and 31 ripples at the top and bottom.
    ($\lambda_1\!=1.29$, $\lambda_2\!=1.33$).} \label{fig:climb}
  \end{center}
  \end{minipage}
\end{figure}
\noindent $j$-flow is induced when the ripples
are not aligned perpendicular  to the main flow, and the simplest
choice for such flow between ripple $j$ and $j+1$ is $ -\Delta m_{lj}
= C_x (x_{j+1} - x_{j})$. The coupling in the $j$-direction is thus
diffusive, and basically acts to align the ripple crests perpendicular
to the oscillatory motion. In the simulations which are presented, the
value of $C_x$ has been fixed to $0.08$. The qualitative results are
not sensitive to a change of this parameters.

To study the motion of defects in this model, we initiate the system
with two patches of nearby wavelengths $\lambda_1$ and $\lambda_2$
separated by two defects. The motion of these defects depends on the
values of the wavelengths (Fig.~\ref{fig:climb}), and we find that
when these are larger than $1.28\!\pm\!0.02$ the defect climbs in the
direction of the lowest wavelength, otherwise it moves in the
direction of the largest wavelength. Thus in a pattern with many
defects, as encountered during the coarsening process, one expects
that only regions with wavelengths larger than
$\lambda_{def}\!=\!1.28$ will survive. We can therefore expect the
final wavelength $\lambda_{eq}$ in a two-dimensional system to lie
between $\lambda_{def}$ and $ \lambda_{max}$.

To check this we have performed simulations in large systems
%100 rows of $100$ to $200$ ripples 
with initial conditions consisting of unstable ripple patterns of
wavelength $0.5$.  The system rapidly coarsens and evolves to a state
where most of the wavelengths lie in the 1D stable regime
$\lambda_{min}\!<\!\lambda\!<\!\lambda_{max}$.  The dynamics then
slows down dramatically and becomes dominated by defect climbing. In
the final relaxed state of the system the peak of the distribution of
ripples lengths lie between $\lambda_{def}$ and $\lambda_{max}$. Runs
with different values of $\lambda_{max}$ have confirmed that the
equilibrium wavelength in the two-dimensional model is systematically
larger than in the one-dimensional case (see Fig.~\ref{leq1D}).

\section{Discussion and conclusion} 

By focusing on a realistic law for the mass exchange between adjacent
ripples, simple models have been formulated that capture a number of
phenomena observed in real ripples. First of all, our model predicts
the existence of a finite band of stable ripple wavenumbers
$\lambda_{min}\!<\lambda\!<\!\lambda_{max}$, which is consistent with
the hysteresis observed in experiments \cite{lofq:80,tray:99}. The
model predicts that the instabilities encountered once these
boundaries are crossed are of short wavelength nature, in agreement
with experiments on one- and two-dimensional sand patterns performed
in Copenhagen \cite{ourexp}.

Coarsening which occurs in the intermediate stages of the development
of vortex ripples is present in our models, and we predict that even
though there is a finite band of stable ripple patterns, the dynamics
selects a well defined final wave length. The exact value of the final
wave length depends on the details of the mass exchange function,
however for a function similar to the results from our simulations of
the fluid and sand flow, we find an equilibrium wavelength of
$\lambda\!=\!1.28$.

The good collapse of the mass exchange function with the maximum shear
stress $\theta_{max}$ indicates that the final wavenumber should be
independent of $\theta_{max}$ as long as suspension is not important.

Following the picture of the mass exchange as in the models, it is
clear that the maximum value of the mass exchange function sets a time
scale for the evolution of the ripples. We have shown that this
maximum value is approximately proportional to $\theta_{max}^{1.5}$.
Thus the time scale of the evolution of the ripples can be expected to
scale as $\theta_{max}^{1.5}$.

Finally we have demonstrated that the models can be applied to two
dimensional ripples patterns, and have found that defect motion
renders the final wavelength of ripples in two-dimensions larger than
in one dimension.

All these predictions are open to experimental verification. In
particular, we are eager to see how consistent the mass exchange
mechanism proposed here is for real data of ripple evolution.

\section{Acknowledgments}
It is a pleasure to acknowledge discussions with Markus Abel, Tomas
Bohr, J{\o}rgen Freds{\o}e, Jonas Lundbek Hansen, Nigel Marsh and
Alexandre Stegner.  M.v.H.~acknowledges financial support from the EU
under contract ERBFMBICT 972554.  M.-L.C.~thanks the Niels Bohr
Institute for hospitality.

\bibliographystyle{plain}

\begin{thebibliography}{1}
\bibitem{notep2} There is some experimental evidence that
in some parameter regime, possibly due to next--nearest--neighbor
interactions, such period-two states can become stable \cite{Scherer}.
\end{thebibliography}

\end{multicols}

\end{document}